\documentclass[reprint,pre,amssymb,amsmath,superscriptaddress,aps,floatfix]{revtex4-1}
\usepackage{amsmath,amsthm,amssymb,amsfonts}
\usepackage{bm}%
\usepackage{graphicx,color}
\usepackage{epstopdf}
\usepackage{siunitx}
\usepackage{hyperref}%
\usepackage{xfrac}
\usepackage{xcolor}
\usepackage{ulem}
\usepackage[T1]{fontenc}
\usepackage{lineno}
\usepackage{multirow}
\usepackage{subcaption}
\expandafter\ifx\csname package@font\endcsname\relax\else
 \expandafter\expandafter
 \expandafter\usepackage
 \expandafter\expandafter
 \expandafter{\csname package@font\endcsname}%
\fi
\newcommand{\ra}[1]{\renewcommand{\arraystretch}{#1}}

\begin{document}
\title{Migration and accumulation of bacteria with chemotaxis and chemokinesis}

\author{Theresa Jakuszeit}
\email{theresa.jakuszeit@curie.fr}
\affiliation{Cavendish Laboratory, University of Cambridge, Cambridge CB3 0HE, U.K.}
\author{James Lindsey-Jones}
\affiliation{Cavendish Laboratory, University of Cambridge, Cambridge CB3 0HE, U.K.}
\author{Fran\c cois J. Peaudecerf}
\affiliation{{Institute of Environmental Engineering, Department of Civil, Environmental and Geomatic Engineering, ETH Z\"{u}rich,
8093 Z\"{u}rich, Switzerland}}
\author{Ottavio A. Croze}
\email{o.croze@physics.org}
\affiliation{Cavendish Laboratory, University of Cambridge, Cambridge CB3 0HE, U.K.}
\date{\today}

\begin{abstract}
Bacteria can chemotactically migrate up attractant gradients by controlling run-and-tumble motility patterns.
In addition to this well-known chemotactic behaviour, several soil and marine bacterial species perform chemokinesis: 
they adjust their swimming speed according to the local concentration of chemoeffector, with higher speed at higher concentration. A field of attractant then induces a spatially varying swimming speed, which results in a drift towards lower attractant concentrations -- contrary to the drift created by chemotaxis. Here, to explore the biological benefits of chemokinesis and investigate its impact on the chemotactic response, we extend a Keller-Segel-type model to include chemokinesis. We apply the model to predict the dynamics of bacterial populations capable of chemokinesis and chemotaxis in chemoeffector fields inspired by microfluidic and agar plate migration assays. We find that chemokinesis combined with chemotaxis not only may enhance the population response with respect to pure chemotaxis, but also modifies it qualitatively. We conclude presenting predictions for bacteria around dynamic finite-size nutrient sources, simulating, e.g., a marine particle or a root. We show that chemokinesis can reduce the measuring bias that is created by a decaying attractant gradient.
\end{abstract}

\flushbottom
\maketitle

\section{Introduction}
\begin{figure}
\includegraphics[width=\columnwidth]{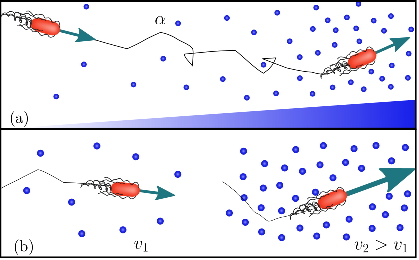}
\caption{Chemotaxis vs. chemokinesis: (a) Chemotaxis is the biased movement of bacteria up a chemical gradient of attractant by reducing the tumbling rate $\alpha$ and, thereby, increasing the length of runs in a favourable direction. (b) Positive chemokinesis leads to an increase in swimming speed $v$ in response to an increase in the local attractant concentration.}
\label{fig:illustration}
\end{figure}

Many bacteria are able to swim by rotating helical flagella distributed on their cell body, and control their swimming pattern by modulating the speed or direction of rotation of their flagellar motors. For example, in the model run-and-tumble motion of \textit{Escherichia coli} \cite{Berg1993}, a bacterium swims approximately straight in a `run' by rotating its flagella in a bundle. When some flagella change their rotation direction, the bundle comes apart and the cell randomly changes direction in a `tumble'. In the absence of external bias, this microscopic pattern resembles a random walk, and leads to macroscopic diffusion of a bacterial population. In the presence of a chemical gradient, the random walk is biased, a response known as chemotaxis. As illustrated in Fig.~\ref{fig:illustration}(a), a bacterium achieves the biased motion up a gradient of attractant by varying the frequency of tumbles in its random walk: if the bacterium moves up the gradient, the tumbling rate $\alpha$ decreases and, thus, the run length increases, while the swimming speed remains constant. 

The run-and-tumble model was originally introduced for enteric bacteria such as \textit{E. coli} and \textit{Salmonella typhimurium} \cite{Berg2004, Tindall2008}, which commonly live in nutrient-rich environments, such as the gut. Marine and soil bacteria, however, often experience heterogeneous and nutrient-scarce environments, and have been found to display different motility patterns. 
For example, several species living in these harsher environments respond to higher concentrations of attractant by increasing their speed \cite{Barbara2003,Garren2014,Armitage1997}. This response, known as `chemokinesis', modifies the swimming speed in response to the local chemical concentration without affecting the tumbling rate, as illustrated in Fig~\ref{fig:illustration}(b). A positive chemokinetic response leads to a higher swimming speed at higher attractant concentrations, whereas a negative response lowers the speed at those concentrations. The strength of positive chemokinetic response can be defined as relative increase in swimming speed over the speed in the absence of chemokinetic effector. A wide range of the chemokinetic response strength has been reported, even for a single species. The responses have been found to vary for the symbiotic soil bacteria \textit{Sinorhizobium meliloti} and \textit{Azospirillum basilense} from $7.5$ to $35\%$ \cite{Attmannspacher2005, Meier2007, Sourjik1996} and $40$ to $77\%$ \cite{Zhulin1993}, respectively; $7.5$ to $73\%$ for the soil and freshwater purple bacterium \textit{Rhodobacter sphaeroides} \cite{Brown1993,Packer1994}; $26$ to $53\%$ for the enterobacterium \textit{E.coli} \cite{Deepika2015}; $48\%$ \cite{Garren2014} or $6$ to $64\%$ \cite{Garren} for the marine pathogen \textit{Vibrio coralliilyticus}. The marine bacterium \textit{V.alginolyticus} showed an increase of up to $80\%$ upon stimulation with glucose in \cite{Son2016}. However, to the best of our knowledge, the pure chemokinetic speed increase as a function of attractant concentration has not been systematically measured for any of these species.

While the role of chemokinesis has been studied extensively in {\it Paramecium} spp. and other protozoa \cite{vanHouten1978,vanHouten1981,vanHouten1982, Giuffre2011}, the biological significance of the chemokinetic response of marine and soil bacteria has yet to be fully elucidated.  
Based on the environment that chemokinetic bacteria have been found in, (positive) chemokinesis might be beneficial in heterogeneous environments with scarce sources of nutrients (attractants). For example, alga-sized microbeads coated with various amino acids were used to study the response of marine bacteria to point-like sources of attractants \cite{Barbara2003}. All marine bacteria studied were observed to accumulate in bands around the point-like sources while displaying a chemokinetic response. Furthermore, chemokinesis could allow marine bacteria to track algae, helping to foster symbioses with these microorganisms, as well as permitting to respond quickly to short bursts of nutrients, such as those generated from lysing algae \cite{Barbara2003a}. Another example of a chemokinetic marine bacterium is the coral pathogen \textit{Vibrio coralliilyticus}. Microfluidic experiments on this pathogenic bacterium in combination with mathematical modelling have suggested that the maximum accumulation in response to chemical cues produced by heat-stressed coral hosts is larger and is reached faster than in the absence of chemokinesis \cite{Garren2014,Garren}. As heat-stressed corals are more susceptible to pathogens, chemokinesis could be a crucial evolutionary advantage in oceans heating up due to climate change. In fact, the chemokinetic response was shown to be even stronger at elevated temperatures; increasing from $6\%$ at $20^\circ$C to $64\%$ at $30^\circ$C \cite{Garren}.

Recent interest in chemokinesis has also been sparked by synthetic microswimmers, such as Janus particles. Janus particles are synthetic colloids in a bath of fuel (e.g. $\rm{H}_2{O}_2$) that propel due to an asymmetric chemical reaction on their surface \cite{Walther}. These particles show a positive chemokinetic response since their swimming speed increases with increasing fuel concentration \cite{Howse2007}, and therefore accumulate in areas of lower fuel concentration.

To date, theoretical work on the combination of positive chemokinesis and chemotaxis has focussed on single-cell level using agent-based models \cite{Garren2014,Son2016,Jackson1987}. The chemokinetic response of the marine pathogen \textit{V. coralliilyticus} has been modelled as a step increase in swimming speed beyond a threshold attractant concentration \cite{Garren2014}. This model was used to analyse the chemotactic response to a transient attractant gradient in a  microfluidic device after an initial release of attractant. The chemotactic index (i.e. the enhancement in cell concentration over a control region) suggests that chemokinesis enables a stronger and faster response. This model has been further adapted to include speed dependent changes in the probability of flicking and reorientation frequency of \textit{V. alginolyticus} \cite{Son2016,Son2013}. In this particular case, the speed induced changes in motility pattern are responsible for a significant part of the chemotaxis improvement as shown by the agent-based model in \cite{Son2016}. 

In this work, we use a continuum model to study the spatio-temporal dynamics of bacterial populations with chemokinetic and chemotactic responses. We incorporate chemokinesis into the standard Keller-Segel model for chemotaxis by deriving the model from microscopic run and tumble dynamics. The model is then used to obtain analytical conditions for chemokinetic drift, and solved numerically for three different example attractant distributions that are inspired by existing experimental systems. 

\section{Model}

We derive a model for chemotaxis in combination with positive chemokinesis by considering a one-dimensional system in which a cell can move either to the right or left with speed $v$. In this system, a right (left) moving particle changes direction with rate $\alpha_R$ ($\alpha_L$). Following previous approaches \cite{Schnitzer1993,Othmer2002}, the one-particle probability density for bacteria evolves according to 
$\partial b/\partial t=-\partial J/\partial x$, and the bacterial flux $J$ can be derived as

\begin{equation}
J(x)=-D_b(x) \dfrac{\partial b}{\partial x} +V(x) b
\label{eq: diffusion-drift}
\end{equation}
\begin{equation}
\centering
\text{with } \quad D_b(x)=\dfrac{v^2}{2\alpha}, \quad V(x)=V_k+V_\chi,
\label{eq:drift velocity}
\end{equation}
where $\alpha_R +\alpha_L=2\alpha$, and where we have defined the chemokinetic
\begin{equation}
V_k=-\dfrac{v}{\alpha} \dfrac{\partial v}{\partial x}
\label{eq:CKdrift}
\end{equation}
and chemotactic
\begin{equation}
V_\chi=\dfrac{v}{2 \alpha} (\alpha_L -\alpha_R),
\label{eq:CTdrift}
\end{equation}
drift speeds, respectively.

We will now connect the bacterial flux (\ref{eq: diffusion-drift}) to the commonly used Keller-Segel model of chemotaxis, adapted to the description of chemokinetic populations in dynamic environments. 


\subsection{Chemotaxis}

For chemotactic populations, the drift-diffusion flux (\ref{eq: diffusion-drift}) is coupled to a chemoattractant density field $c$ via the chemotactic drift speed, given by Eq. (\ref{eq:CTdrift}). In the standard Keller-Segel model it is phenomenologically asserted that this chemotactic drift speed is proportional to the change in the chemical attractant in space, $V_\chi=\chi \nabla f_\chi$,
where $\chi$ is the chemotactic sensitivity parameter and $f_\chi$ is a function of $c$ that ensures that the chemotactic drift is biased towards higher attractant concentrations \cite{Cates2012}. This definition of the chemotactic drift speed assumes that the chemical attractant profile changes in space but not in time. However, as mentioned, known chemokinetic bacteria inhabit dynamic environments, such as in the ocean or soil. As pointed out by Hein et al \cite{Hein2016}, the effective gradient perceived by a bacterium changes in a temporally varying attractant profile depending on the direction of its run.
Consider a source of attractant at one point of space, and the associated negative gradients of concentration as one gets away from it. If these gradients are steady in time, for example as in a microfluidic setting, a bacterium exploiting the attractant landscape will correctly detect the gradient and swim towards the source, as in standard chemotaxis. However, if the source corresponds to a single point-like release of attractant, and the concentration at the source position also decays due to diffusion, a bacterium travelling toward the source perceives a smaller increase (or even a decrease) in concentration compared to the steady case. On the other hand, a bacterium moving away from the source perceives a decrease, which is reduced in magnitude compared to the steady case.
We show in Appendix \ref{app:ChemotacticSensitivity} how this influences the mean run length of a bacterium, and, in light of this, modify the chemotactic drift speed to

\begin{equation}
V_\chi=\chi \left( \nabla f_\chi + \dfrac{1}{v} \partial_t f_\chi \right),
\label{eq:chemotactic drift velocity Updated}
\end{equation}
where we see that the presence of a temporally increasing (decreasing) gradient increases (decreases) the chemotactic drift speed. Further, we note that the perturbation to the chemotactic drift speed is smaller the larger the value of the swimming speed. For the chemotactic function $f_\chi$, we may choose, for example \cite{Croze2011}, 
\begin{equation}
f_\chi=\dfrac{c(x,t)}{c(x,t)+k_\chi}.
\label{eq: chemotactic bias}
\end{equation}

Following \cite{DeGennes2004}, using relationship (\ref{eq:CTdrift}) for the chemotactic speed, the microscopic swimming parameters $v$ and $\alpha$ can be related to the macroscopic parameter $\chi$ (see Appendix \ref{app:ChemotacticSensitivity} for details) as 
\begin{equation}
\chi=\dfrac{v^2}{\alpha} \beta,
\label{eq: chemotacticSensitivityConstant}
\end{equation}
where $\beta$ is constant dependent on the chemotactic response (again, see Appendix \ref{app:ChemotacticSensitivity} for details). 

\subsection{Chemokinesis}

In this section we consider how chemokinesis can be modelled. We can modify Eq.~\eqref{eq: chemotacticSensitivityConstant} to include chemokinesis (a spatially varying swimming speed) to give
\begin{equation}
\chi(x)= \chi_0 \dfrac{v(x)^2}{v_0^2},
\label{eq: chemotactic sensitivity}
\end{equation}
in which the subscript refers to the parameters in the absence of chemokinesis (the interested reader can find a detailed derivation in Appendix \ref{app:ChemotacticSensitivity}). Here we assumed that the swimming speed is constant during a run. 

In addition to modifying the chemotactic drift $\chi$, the spatially varying swimming speed of chemokinetic bacteria
causes an additional chemokinetic drift to arise with a speed given by Eq~(\ref{eq:CKdrift}).
When acting alone, this drift drives the cells towards regions of lower speed \cite{Schnitzer1993}. 

\begin{figure}
\includegraphics[width=\columnwidth]{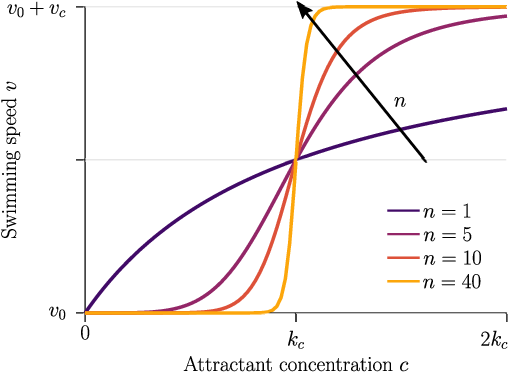}
\caption{Chemokinetic response function. The swimming speed increases from a reference speed, $v_0$, by at most $v_c$, depending on the local concentration of the attractant according to a Hill type function. The parameter $n$ determines the strength of the gradient at the half-saturation concentration, $k_c$.}
\label{fig:ModelIllustration-Speed}
\end{figure}

Next we consider how to quantify the chemokinetic coupling between speed and local concentration of attractant. We need to make assumptions about the relationship between speed and attractant concentration, as experimental studies have not been carried out to provide this. Firstly, we assume that cells swim at a base level speed, $v_0$. Secondly, we reasonably posit that chemokinesis monotonically increases the swimming speed up to a maximum speed denoted as $v_0 + v_c$. The dynamics between these limits are given by an unknown function characterising the chemokinetic response. In the following, we choose a Hill-type equation
\begin{equation}
v(x)=v_0+v_c \dfrac{c(x)^n}{c(x)^n+k_c^n}
\label{eq: dim speed function}
\end{equation} 
to approximate the chemokinetic response. The Hill parameter $n$ allows us to introduce an inflection point and change the gradient $\partial v/\partial c$. As can be seen from Fig~\ref{fig:ModelIllustration-Speed}, the speed increases monotonically with increasing attractant concentrations $c$ for all $n$ and $v_c>0$, where the half-maximum speed is reached at the attractant concentration $k_c$. For any $n$, $v=v_0+k_c/2$ at $c=k_c$ as all functions covered by Eq. \eqref{eq: dim speed function} have the same half-saturation constant. Note that upon setting $n=1$, one recovers a Michaelis-Menten type response; for $n\rightarrow \infty$, Eq.~\eqref{eq: dim speed function} approaches a step function, which has been used previously to approximate the chemokinetic response \cite{Garren2014, Son2016}. 

Finally, in situations where bacterial growth cannot be ignored (e.g. migration across agar plates considered in section \ref{sect:agar-plate}), bacteria are also assumed to undergo logistic growth, which comprises growth and death terms \cite{Croze2011}. We note, however, that the chemoeffectors eliciting chemotaxis and chemokinesis need not in general be metabolisable nutrients that induce growth. The full model equations for the chemical attractant field and the chemotactic and chemokinetic bacterial population field, therefore, are

\begin{subequations}\label{eq: dim model}
\begin{align}
\dfrac{\partial c}{\partial t}&=D_c \dfrac{\partial^2 c}{\partial x^2}-k_g g(c) \dfrac{b}{Y} \label{eq: dim model chemical}\\
\dfrac{\partial b}{\partial t}&= -\dfrac{\partial J}{\partial x}+k_g  g(c) b \left(1 - \dfrac{b}{k_b} \right)  \label{eq: dim model-g bacteria}\\
J(x)&= -\dfrac{v^2}{\alpha} \dfrac{\partial b}{\partial x}  - \dfrac{v}{\alpha} \dfrac{\partial v}{\partial x} b  +
\chi(x) \left( \nabla f_\chi + \dfrac{1}{v} \partial_t f_\chi \right)b, \label{eq: dim model flux}
\end{align}
\end{subequations}
where $k_g$ is the maximum growth rate, $Y$ is the bacterial yield, and $k_b$ is the carrying capacity. Furthermore, $g(c)$ is chosen as a Monod-type growth function $g(c)=c/(c+k_s)$ with the half-saturation constant $k_s$. Note that $\chi(x)$ is given by Eq. \eqref{eq: chemotactic sensitivity}, $f_\chi$ by Eq. \eqref{eq: chemotactic bias} and $v(x)$ by Eq. \eqref{eq: dim speed function}.



%

Let us summarise the effects of chemokinesis in our model. A spatially varying speed affects all three terms of the bacterial flux $J$ in Eq.~\eqref{eq: dim model flux} as: (i) there are regions with a higher diffusivity since $D_b \sim v^2$ (first term); (ii) it introduces a drift where $\partial v/\partial x \neq 0$ (second term); and (iii) there are regions with a larger chemotactic drift as $\chi \sim v^2$ (third term).

\subsection{Non-dimensionalisation}

The system of partial differential equations \eqref{eq: dim model} is non-dimensionalised using the characteristic time and length scales $t_0=k_g^{-1}$ and $x_0=\sqrt{t_0 D^0_b}$, where the bacterial diffusivity is $D^0_b=v_0^2\alpha^{-1}$. We rescale the attractant and bacterial densities by their respective initial densities, $c_0$ and $b_0$. The system of PDEs in dimensionless form thus reads

\begin{subequations} \label{eq: nondim model}
\begin{align}
\dfrac{\partial C}{\partial T}&= N \nabla^2 C - HB \dfrac{C}{C+K_S} \label{eq: nondim model chemical}\\
\dfrac{\partial B}{\partial T}&= -\nabla J + B \dfrac{C}{C+K_S} \left( 1 - B \right) \label{eq: nondim model bacteria} \\
J&=-\mathcal{V}(X)^2  \nabla B \: + V_k B \: + V_\chi B \label{eq: nondim model flux} \\
\mathsf{V}_k&=-\mathcal{V}(X) \dfrac{n \eta \omega^n C^{n-1}}{(C^n + \omega^n)^2}\nabla C \label{eq: nondim model chemokinetic drift}\\
\mathsf{V}_\chi&=\mathcal{V}(X)^2 \dfrac{\delta_0 K_\chi}{(C+K_\chi)^2} \left(\nabla C + \dfrac{\zeta}{\mathcal{V}(X)} \dfrac{\partial C}{\partial T} \right) \label{eq: nondim model chemotactic drift}
\end{align}
\end{subequations}
with non-dimensional parameters $B=b/b_0$, $C=c/c_0$, $N=D_c/D^0_b$, $H=b_0/(Y c_0)$, $K_S=k_s/c_0$, $\eta=v_c/v_0$, $\omega=k_c/c_0$, $\delta_0=\chi_0/D_b^0$, $\zeta=v_0 /(\alpha x_0)$ and $K_\chi=k_\chi/c_0$, and non-dimensional time $T=t/t_0$ and space $X=x/x_0$.
The non-dimensional speed function in Eq.~\eqref{eq: nondim model} is

\begin{equation}
\mathcal{V}(X)=1+\eta \dfrac{C^n}{C^n + \omega^n},
\label{eq: nondim speed function}
\end{equation}
where $\eta$ is the maximum increase in swimming speed, $\omega$ is the attractant concentration at which the half maximum speed increase is reached, and $n$ is the Hill parameter. Note that $\omega$ corresponds to the inflection point of Eq.~\eqref{eq: nondim speed function}. The chemokinetic response is positive for $\eta>0$.

Finally, the model is extended to the 2D axisymmetric case by introducing polar coordinates as 

\begin{subequations}
\begin{align}
\nabla B&=\dfrac{\partial B}{\partial R} \\
\nabla^2 B &= \dfrac{\partial^2 B}{\partial R^2} + \dfrac{1}{R} \dfrac{\partial B}{\partial R},
\end{align}
\end{subequations}

and the equivalent equations for the chemical field, $C$.
The details of the numerical solution and simulations of the model are described in appendix \ref{app: NumericalSolution}, including a summary of parameters used. The parameters for the results presented in the main text were chosen to best illustrate the chemokinetic effect (see Appendix B for a discussion of parameter values and the Supplementary Information for further simulations with different parameter sets, including $n$). In section \ref{sec:Results} we will present the simulation results for three different types of attractant gradient. 

\subsection{Condition for dominant chemokinetic drift}
\label{sec:DriftCondition}

Before solving the extended Keller-Segel model numerically, we can use Eq.~\eqref{eq: nondim model} to analytically derive a condition on the relative importance of chemokinetic and chemotactic contributions to the drift of the bacterial population. The drift due to a spatially varying swimming speed causes cells to accumulate in regions where they have low speeds. For $\eta>0$, by construction of the velocity function \eqref{eq: nondim speed function}, the speed is low at low attractant concentrations. 
The chemotactic drift, on the other hand, is directed towards higher attractant concentrations by virtue of Eq.~\eqref{eq: chemotactic bias}. Hence, the bacterial density is governed by two competing drifts, as can be seen from the opposing signs in Eq.~\eqref{eq: nondim model chemokinetic drift} and \eqref{eq: nondim model chemotactic drift}. If the chemokinetic drift is larger than the chemotactic drift for a large part of the spatial domain, this could lead to accumulation at low attractant concentrations, instead of the biologically-desirable accumulation at high concentrations. Assuming a stationary and linear attractant profile (i.e. $\partial C/\partial T=0$ and $\partial C/\partial X=\mathrm{const}$), we have from Eq.~\eqref{eq: nondim model chemokinetic drift} and \eqref{eq: nondim model chemotactic drift} that the chemokinetic drift is larger than the chemotactic drift if

\begin{equation}
\dfrac{n \eta \omega^n}{(C^n+\omega^n)^2} \mathcal{V}^{-1} C^{n-1} > \dfrac{\delta_0 K_\chi}{(C+K_\chi)^2} \quad \forall X \in \Omega,
\label{eq: condition1}
\end{equation}

where $\Omega$ is the spatial domain. In the case of a linear attractant profile, we know from Eq.~\eqref{eq: nondim speed function} (and Fig.~\ref{fig:ModelIllustration-Speed}) that the gradient $\partial V/\partial C$ is maximum close to the half-saturation constant $\omega$. Thus, we evaluate condition \eqref{eq: condition1} with $C=\omega$, which yields the condition 
\begin{equation}
n > \dfrac{4 \delta_0 \omega K_\chi}{(\omega + K_\chi)^2} \left( \dfrac{1}{\eta} + \dfrac{1}{2} \right) \forall X \in \Omega: C(X)=\omega .
\label{eq: conditionFull}
\end{equation}
If the Hill parameter $n$ exceeds this threshold, the chemokinetic drift is predicted to be larger than the chemotactic drift at the attractant concentration $C^*=\omega$ for $\eta>0$. Note that, for a step function, condition \eqref{eq: conditionFull} is always met at the threshold concentration $C^*=\omega$ since $n \rightarrow \infty$. Conversely, if $\eta<0$ (i.e. modelling a negative chemokinetic response), condition \eqref{eq: conditionFull} will never be met as the chemotactic and chemokinetic drift have the same direction (chemokinesis in this case is stabilising). 

\section{Results}
\label{sec:Results}

\subsection{Steady linear attractant profile}

\begin{figure}
\includegraphics[width=\columnwidth]{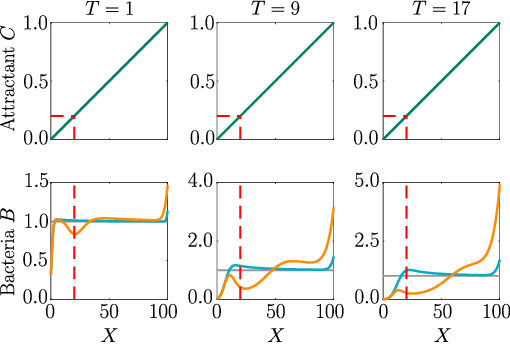}
\caption{Chemokinesis in steady linear attractant profile. Bacterial response (bottom row) to a fixed linear attractant profile (top row) using pure chemotaxis (blue curve) versus chemotaxis with chemokinesis (orange curve) for three time points. Note the changing range of the y-axis in the bottom row. The initial bacterial profile is indicated by the grey line. The position at which $C=\omega$ is highlighted by a dotted red line. Parameters $H=0, \, N=0, \, K_\chi=0.53, \, \delta_0=50,\, \eta=2, \, \omega=0.2, \, n=5, \, T=1,9,17$.}
\label{fig:SteadyGradient}
\end{figure}

The full model Eq.~\eqref{eq: nondim model} includes the effect of growth and consumption as well as chemotaxis and chemokinesis. Thus, any change in the spatial distribution of the bacterial population due to chemotaxis and chemokinesis will feed back onto the attractant distribution due to consumption by the bacteria. In order to identify the influence of chemokinesis without such additional complications, we first solve the model for a steady attractant gradient, i.e. $\partial C/ \partial T=0$. We furthermore assume a linear profile such that $\partial C / \partial X=\mathrm{const}$ and ignore consumption, and thus population growth. This situation might be achieved experimentally in microfluidic devices \cite{Rusconi2014}, where the gradient may be fixed and bacterial growth can be neglected on experimental timescales short compared to growth timescales $\sim k_g^{-1}$. 

\begin{figure}
\includegraphics[width=\columnwidth]{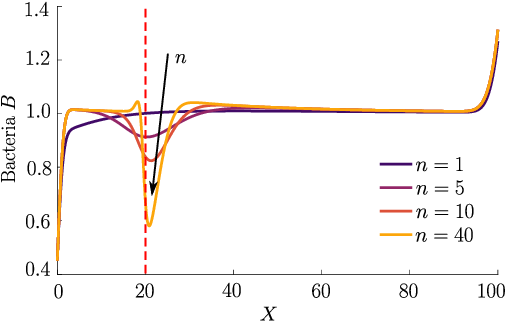}
\caption{Effect of Hill parameter on chemokinesis in steady linear attractant profile. Chemotactic-chemokinetic bacterial response to a fixed linear attractant profile at $T=0.5$ (same constant attractant profile as in Fig.~\ref{fig:SteadyGradient}). The Hill parameter $n$ in Eq.~\eqref{eq: nondim speed function} and, thus, the speed gradient is varied. The position at which $C=\omega$ is highlighted by a dotted red line. Parameters: $H=0, \, N=0, \, K_\chi=0.53, \, \delta_0=50,\, \eta=2, \, \omega=0.2, \, n=1,5,10,40$. For the chosen parameters, $n=40$ is just above the threshold given by \eqref{eq: conditionFull}.}
\label{fig:SteadyGradientHill}
\end{figure}

Fig.~\ref{fig:SteadyGradient} compares the response of a purely chemotactic population to the response of a chemotactic-chemokinetic population. Chemokinesis leads to a stronger and faster accumulation than in the purely chemotactic case. At the critical concentration, $C=\omega$, however, the chemokinetic drift holds back a subset of the population because it is directed towards lower attractant concentrations as described in section \ref{sec:DriftCondition}. This can be seen in the form of an accumulation of cells at low attractant concentrations. As the chemotactic sensitivity parameter, $\delta_0$, is large in this simulation, the population subset overcomes the drift and accumulates at high attractant concentrations at long times. However, if condition \eqref{eq: conditionFull} is fulfilled, the chemokinetic drift is larger than the chemotactic drift. Thus, there is a subpopulation driven to small attractant concentrations by chemokinesis. The effect of varying the Hill parameter $n$ in Eq.~\eqref{eq: nondim speed function} is illustrated in Fig.~\ref{fig:SteadyGradientHill}. For the parameters chosen in Figure \ref{fig:SteadyGradient}, condition \eqref{eq: conditionFull} is met for $n>39.78$. In experiments, this would require observing the transient bacterial concentration profiles in addition to the commonly reported steady-state profiles.

As the attractant concentration is fixed, we can determine a steady-state for the bacterial population, i.e. by setting $\partial B/\partial T=\partial J/\partial X=0$. Due to the homogeneous Neumann boundary conditions of the problem, we have $J=0$ in Eq.~\eqref{eq: nondim model}, which yields the differential equation

\begin{equation}
\dfrac{\partial B}{\partial X} + B\left[ \dfrac{1}{2D}\dfrac{\partial D}{\partial X} - \delta_0 \dfrac{d}{dX}\left(\dfrac{C}{C+K_\chi}\right)\right]=0,
\end{equation}

where we used $D(X)=\mathcal{V}(X)^2$. This equation can be integrated to give the steady-state

\begin{equation}
\dfrac{B}{B^*}=\dfrac{\mathcal{V}^*}{\mathcal{V}}\exp\left\lbrace \delta_0 \left(\dfrac{C}{C+K_\chi}-\dfrac{C^*}{C^*+K_\chi}\right) \right\rbrace,
\label{eq: steadyStateSolution}
\end{equation} 
where $B^*,C^*,V^*$ are reference values at a chosen reference point $X^*$. It is thus clear that, in addition to the influence on the dynamics, chemokinesis affects the steady-state solution via the term $\mathcal{V}^*/\mathcal{V}$, where $\mathcal{V}$ varies in space due to chemokinesis. In the case of chemokinesis but no chemotaxis (i.e. $\delta_0=0$), the steady-state is determined by the inverse of the speed distribution, i.e. the bacteria accumulate at low speed, as expected and shown previously \citep{Schnitzer1993,Othmer2013, Arlt}. For non-zero $\delta_0$, if the speed is uniform in space, $\mathcal{V}^*/\mathcal{V}=1$ and Eq.~\eqref{eq: steadyStateSolution} reduces to the chemotactic steady-state solution. The exponential term in Eq.~\eqref{eq: steadyStateSolution} represents the chemotactic contribution to the steady state, which does not depend on the swimming speed. Thus, the increase in chemotactic sensitivity $\chi$ (see Eq.~\eqref{eq: chemotactic sensitivity}) must be balanced by the increase in diffusivity at steady state in a fixed chemical gradient. However, chemokinesis still affects the steady state via the term $\mathcal{V}^*/\mathcal{V}$. This chemokinetic effect in the steady-state may only be detectable in experiments with small $\delta_0=\chi_0/D_b^0$, since the chemotactic exponential term will dominate $\mathcal{V}^*/\mathcal{V}$ for large $\delta_0$. 


\subsection{Self-generated gradient: agar plate \label{sect:agar-plate}}

\begin{figure}
\includegraphics[width=\columnwidth]{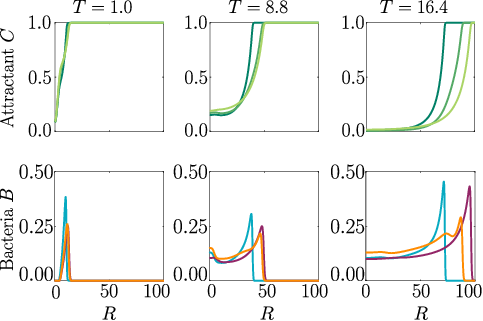}
\caption{Self-generated gradient. Bacterial populations (bottom row) create an attractant gradient (top row) via consumption, which they respond to with chemotaxis at base speed $\mathcal{V}=1$ (blue curve) or chemotaxis-chemokinesis (orange curve). Chemokinesis leads to a faster but also broader, less pronounced bacterial wave. The chemotactic population travelling at constant speed $\mathcal{V}=1+\eta$ (purple curve) has the fastest travelling pulse. Parameters $H=3.5, \, K_S=1, \, N=0.5, \, K_\chi=0.53, \, \delta_0=105,\, \eta=0.5, \, \omega=0.5, \, n=5, \, T=1,8.8,16.4$}
\label{fig:AgarPlate}
\end{figure}

We now consider an evolving attractant field with consumption and growth of bacteria. The attractant is initially uniformly distributed in a 2D axisymmetric setting. This set-up is reminiscent of the classical agar plate experiments, in which bacteria are inoculated in the centre of a nutrient agar plate, see e.g. \cite{Adler66, Croze2011}. While growing and consuming nutrient, the population creates a gradient of attractant, which it then follows outwards in a chemotactic wave. The attractant profile is a travelling wave itself, and we assume here that the profile relative to the bacterial travelling wave is stationary, i.e. $\zeta=0$ in the chemotactic drift \eqref{eq: nondim model chemotactic drift}.

In Fig.~\ref{fig:AgarPlate}, we compare two chemotactic populations to a chemotactic-chemokinetic population. The chemotactic populations travel at a constant speed, either $\mathcal{V}=1$ (blue curve) or $\mathcal{V}=1+\eta$ (purple curve). Both populations develop a sharp travelling wave, with a larger wave speed for the population at speed $\mathcal{V}=1 +\eta$. The chemotactic-chemokinetic bacterial population, on the other hand, develops a broader wave profile. The peak of the wave front is smaller and is followed by a plateau. This effect is more pronounced at late times, as can be seen in the third lower panel of Fig.~\ref{fig:AgarPlate}. The reduced pulse also travels slower than the pulse of the chemotactic population at elevated speed $\mathcal{V}=1+\eta$ because the front speed scales with the number of bacteria in the pulse \cite{Seyrich2019}. 
This observation might explain why in agar plate experiments testing for chemotaxis, chemokinetic species such as \textit{Sinorhizobium meliloti} lack the sharp bands \cite{Platzer1997,Meier2007}, which are known to be a hallmark of chemotaxis for other species, e.g. \textit{E. coli} \cite{Adler66, Croze2011}.

\begin{figure}
\includegraphics[width=\columnwidth]{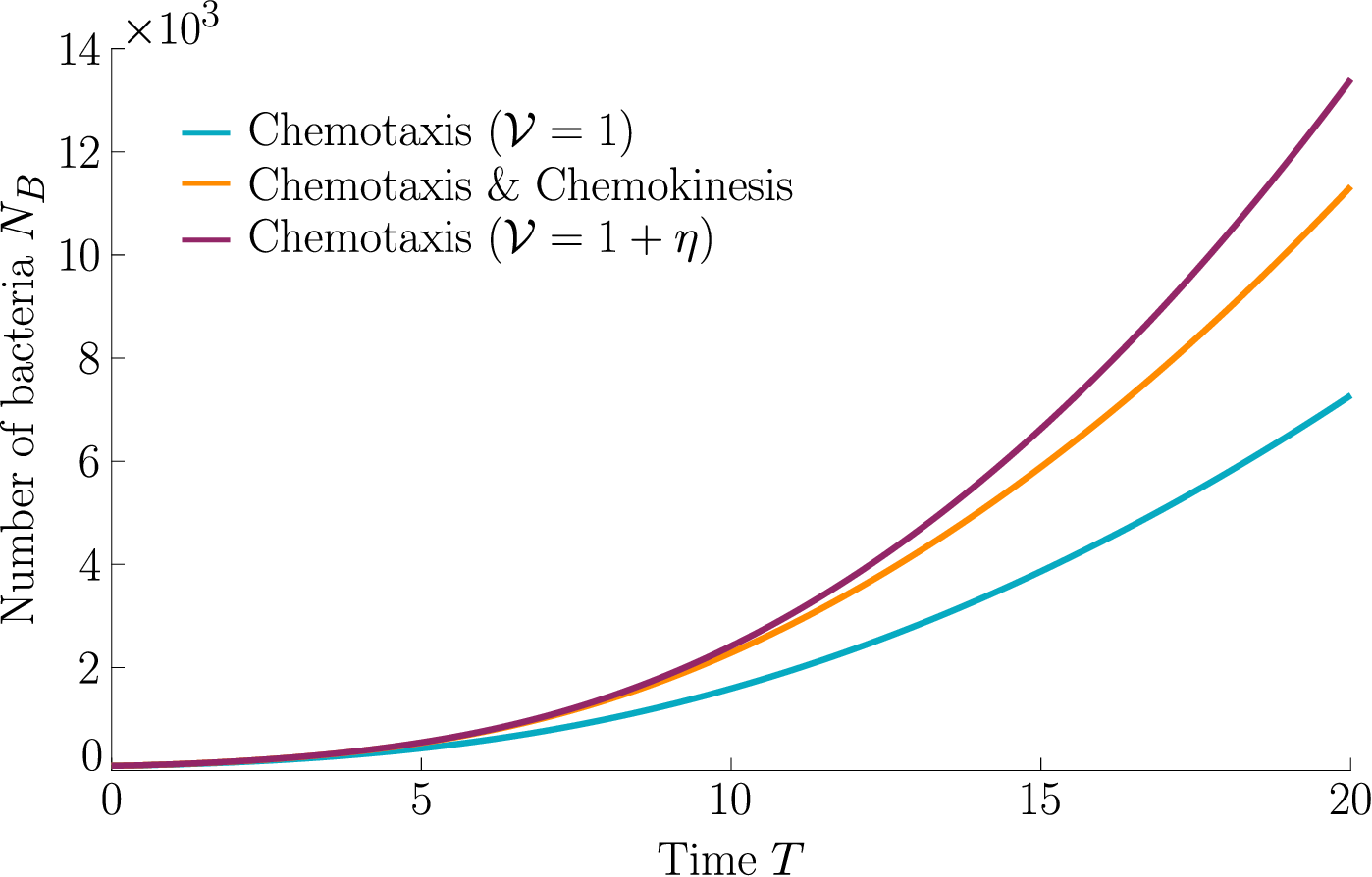}
\caption{Population growth. The size of the populations in Fig. \ref{fig:AgarPlate} is the bacterial density integrated over the simulation domain. Faster travelling waves of the chemokinetic-chemotactic population (orange curve) and the chemotactic population at constant speed $\mathcal{V}=1+\eta$ (purple curve) also induce a faster population growth due to consumption of nutrients, compared to the chemotactic population travelling at $\mathcal{V}=1$ (blue curve). Parameters as in Fig. \ref{fig:AgarPlate}}
\label{fig:AgarPlate_integration}
\end{figure}

Chemokinesis confers an additional biological advantage in the form of increased population growth as can be seen in Fig. \ref{fig:AgarPlate_integration}, which shows the integrated number of cells over time. At any point in time, the chemotactic population is smaller than the chemotactic-chemokinetic population. However, the chemotactic population at elevated speed, $\mathcal{V}=1+\eta$, (purple curve) shows a stronger population growth than the chemotactic-chemokinetic population due to the faster travelling pulse, which is caused by the increased swimming speed. This increase in swimming speed is associated with metabolic cost \cite{Martinez2014}. Thus, permanently swimming faster independent of the attractant concentration could be a beneficial strategy, if the metabolically available energy is not constrained by nutrient supply. 
When nutrient concentrations are low, on the other hand, increasing swimming speed provides no benefit to bacteria and metabolism is a limiting factor. Chemokinesis could provide an advantageous speed enhancement when it is both metabolically affordable and beneficial \cite{Barbara2003, Barbara2003a}. While the situation considered in this section assumed an abundant supply of chemoeffectors (and in this case nutrients), the next section will consider the response to a transient burst of chemoeffectors.

\subsection{Transient source\label{sect-transource}}

A localised burst of chemoeffector may, e.g., occur in the sea if algae/phytoplankton lyse and release their content, as has been recently studied in the laboratory \cite{Smriga2016}, or when marine particles exude plumes of chemoeffector \cite{Stocker4209}. In soil, plant roots exude sugars and other potential nutrients, which locally create a high concentration of chemical attractants. In the following, we consider a single strong pulse of chemoeffector originating from a finite-size axisymmetric source that dissipates via diffusion, modelling a potential dynamic environment around roots or marine particles. The attractant profile that develops is $C(R,T)=S(4\pi N T)^{-1} \exp(-R^2/4 N T)$, with $S$ representing the amount of chemoeffector contained in the pulse in non-dimensional units. The attractant profile is illustrated in Fig. \ref{fig:PointSource} (top row), while the bacteria are initially uniformly distributed in the domain at concentration $0.2$. To model the response to such a transient attractant profile, we need to include the chemotactic drift velocity, Eq.~\eqref{eq: nondim model chemotactic drift} with $\zeta>0$, modified to account for the effective gradient perceived by bacteria as they traverse the temporally varying pulse (see Appendix \ref{app:ChemotacticSensitivity}). 

\begin{figure}
\includegraphics[width=\columnwidth]{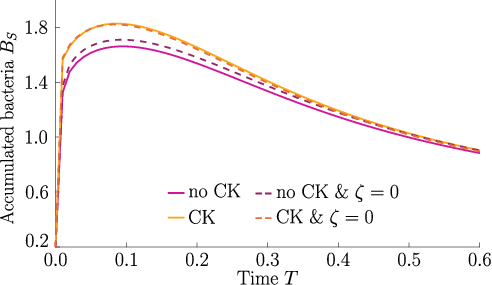}
\caption{Accumulation at a transient source. The bacterial accumulation at the source is reduced due to the reduced drift $V_\chi$ if the transient nature of the attractant profile is taken into account (i.e. $\zeta \neq 0$). Chemokinesis (CK) mitigates for this effect. Parameters $H=3.5, \, K_S=1, \, N=0.5, \, K_\chi=0.53, \, \delta_0=50,\, \zeta=8.164\cdot 10^{-3},\, \eta=2$ ($\eta=0$ if no CK), $\, \omega=0.2, \, n=1, \, S=0.5$; no bacterial growth. If $V_\chi$ includes the effect of transient source, $\zeta=8.164\cdot 10^{-3}$, otherwise $\zeta=0$.}
\label{fig:PointSourceMaximum}
\end{figure}

As can be seen from Fig. \ref{fig:PointSource}, bacteria with chemotaxis and chemokinesis display a faster and stronger response to a chemoeffector pulse than those with chemotaxis alone. This strong accumulation occurs in spite of the fact that, because of chemokinesis, diffusivity close to the source is higher for these bacteria. For the transient pulse under consideration, temporal variations in the chemoeffector concentration ($\sim \partial_t f_\chi$) need to be considered when modelling the chemotactic response. Indeed, the run duration for a bacterium travelling up/down the gradient is $\tau^{L,R} \propto \pm (v\nabla f_\chi + \partial_t f_\chi)$ (see Appendix \ref{app:ChemotacticSensitivity}). Thus, when $v$ is higher, as for chemokinetic bacteria, the effect of temporal variation on the chemotactic bias of tumbles is reduced, increasing the accuracy of the chemotactic response. This effect contributes to accounting for the stronger accumulation of chemotactic-chemokinetic bacteria. To illustrate this further, in Fig. \ref{fig:PointSourceMaximum} we plot the number of bacteria accumulated at the attractant source, $B_S$ (the maximum of the bacterial profiles shown in Fig. \ref{fig:PointSource}), as a function of time, for chemotactic and chemotactic-chemokinetic populations. The plot displays bacterial accumulation when the temporal perturbation to the chemotactic response is included in the model ($\zeta>0$ in \eqref{eq: nondim model chemotactic drift}) and when it is not ($\zeta=0$). In the case of a purely chemotactic population, it can be seen that the predicted amount of bacteria accumulated at the source is lower for a model that ignores the temporal perturbation than for one that includes it. Chemokinesis, on the other hand, reduces the relative effect of temporal perturbation so much that there is very little difference between model predictions with $\zeta=0$ and $\zeta>0$. 

\begin{figure}
\includegraphics[width=\columnwidth]{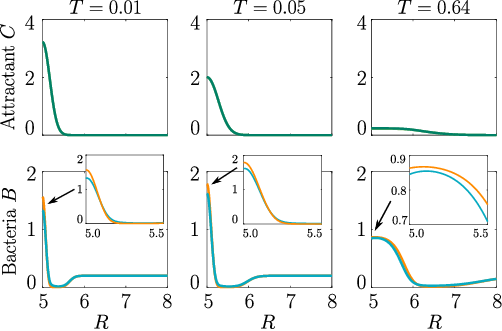}
\caption{Time evolution of the bacterial response to a diffusing attractant from a transient source. Bacterial populations (bottom row) are attracted to source of diffusing attractant (top row). The chemokinetic-chemotactic population (orange curve) shows a faster and stronger accumulation than the purely chemotactic population (blue curve). Parameters $H=3.5, \, K_S=1, \, N=0.5, \, K_\chi=0.53, \, \delta_0=50,\, \eta=2, \, \omega=0.2, \, n=1, \, S=0.5, \, T=0.01,0.05,0.64$; no bacterial growth.}
\label{fig:PointSource}
\end{figure}
 
These results suggest that a chemokinetic population might be able to overtake purely chemotactic competitors in response to a sudden nutrient release. While the difference observed in Fig.\ref{fig:PointSourceMaximum} of at most $\sim 12\%$ may seem small, a corresponding boost to the growth rate can be sufficient to outcompete a purely chemotactic strain within a few generations. 
In such transient nutrient landscapes, we have further shown that chemokinesis can 
reduce the adverse effect that a temporal change in attractant profile can have on the chemotactic response.

\section{Discussion}

Chemokinesis is a known response for many environmentally relevant bacteria, yet its consequences for bacterial population dynamics have been little explored. In this work, using a modified Keller-Segel continuum model we have shown how chemokinesis significantly affects both the dynamics and steady-states of bacterial populations capable of chemotactic and chemokinetic behaviour. Our model incorporates the effects of a concentration dependent speed, including an increase in the chemotactic sensitivity, and a recently suggested modification to the chemotactic response in dynamic environments \cite{Hein2016}, which was derived adapting the microscopic model first suggested by de Gennes \cite{DeGennes2004}.


We have solved our model numerically to explore the effect of chemokinesis on migration and accumulation in experimentally realistic gradients. In a fixed attractant gradient, our results show that chemokinesis can lead to two subpopulations travelling at different speeds, with the slower one being held back by the chemokinetic drift.
In the case of agar plate migration, where bacteria inoculated onto the plate generate their own gradient by consuming nutrients, we find that populations with chemokinesis migrate out from the inoculation point in waves that are faster, but broader than purely chemotactic migrating populations. While the increase in front speed could be explained by a population at a uniformly increased swimming speed, the broadening is only observed in the chemokinetic population. It is a new feature not predicted by previous studies using agent based simulations \cite{Son2016}. It is possible that this broadening might explain why the classic chemotactic  Adler bands observed for \textit{E. coli} \cite{Adler66} are not observed for chemokinetic soil bacteria such as \textit{Sinorhizobium meliloti}  \cite{Platzer1997,Meier2007}. Furthermore, chemokinesis increases the population growth significantly in comparison to purely chemotactic migration. 

Our simulations also considered the case of a transient source of nutrients, e.g., a lysed algal cell. In this case, our results show how bacteria with chemokinesis and chemotaxis accumulate faster and more strongly around the source, while concentrations of nutrients are high, with respect to purely chemotactic bacteria. This chemokinetic advantage is both due to the enhanced migration discussed in the previous examples, but also to the fact that chemokinesis mitigates the perturbation to the chemotactic response due to the transient nature of the attractant profile. We note that, while our model includes the effect of transient chemical fields on tumbles, it does not include a recently modelled effect of chemokinesis on the precision of chemosensing \cite{Brumley10792}. It will be interesting to include this additional effect, which could lead to further enhancements in chemokinetic accumulations, in future elaborations of our model.


The role of metabolism is an important consideration for chemokinesis. For example, positive chemokinesis might be caused purely by an increase in nutrient availability. A resulting increase in the energy level of the cell may lead to more energy being available for flagella rotation, which allows the cell to swim faster. As the chemotactic sensitivity scales with the swimming speed as $\chi \propto v^2$, a faster swimming population will always show a stronger chemotactic drift. However, swimming is associated with a considerable metabolic cost \cite{Martinez2014,Mitchell2002,Ni2020}. Indeed, the energetic cost of swimming increases quadratically with the swimming speed \cite{Taylor2012}. Therefore, it might be beneficial to swim faster (and, thereby, improve chemotaxis) only if favourable nutrient conditions are available.
While the energy level of a cell may influence the extent of the chemokinetic response, there are several indications of metabolism-independent chemokinesis for different bacterial species \cite{Deepika2015,Garren2014,Packer1994}. In a dynamic environment such as considered in section \ref{sect-transource}, chemokinesis can then help to reduce temporal bias and improve the chemotactic response. As pointed out by Hein et al, the threshold for detecting absolute concentration is smaller than for gradient detection \cite{Hein2016}. Thus, chemokinesis can take place at lower background concentrations to improve chemotaxis only when needed. 

The predictions of our model include interesting qualitative effects, which have not previously been observed in agent-based models: the slower subpopulation in the fixed attractant profile, and a broadening of the travelling wave in a self-generated gradient. To test these predictions experimentally, chemotaxis and chemokinesis should be addressed independently. For example,
a recently developed \textit{E.coli} system with a swimming speed that is controlled via light \cite{Arlt} could be used to engineer populations with a swimming speed that can be controlled independently of chemotaxis. 

Future theoretical investigations would benefit significantly from the experimental measurement of the chemokinetic response function relating swimming speed and local attractant concentration. In this work we assumed a smooth change from a reference speed to an increased speed, where the degree of change in swimming speed changed with a single parameter. As we have shown, a very steep change in swimming speed (e.g. in form of a step change as assumed previously in agent-based models \cite{Garren,Son2016}), could actually inhibit chemotaxis rather than promote it. Experimental work so far has been restricted to measurements at very few different attractant concentrations, which makes it difficult to deduce a functional relationship between speed and attractant concentration. Thus, further work is required to determine the function $v(c)$ for chemokinetic bacterial species. Such measurements would also allow elucidation of the rate of adaptation, i.e. how quickly the swimming speed adapts to its new value both for an increase and decrease in attractant concentration, and which we have here assumed instantaneous. 
The increase in the population-averaged swimming speed in response to a uniform addition of effector occurred on the time-scale of $100-200$s for \textit{V. alginolyticus} \cite{Son2016}. In \cite{Packer1994}, on the other hand, the chemokinetic response of \textit{R. sphaeroides} was measured within $10$s upon uniform addition of chemoeffector, at which point the swimming speed had already adapted to its increased level. Furthermore, a desensitisation to a sustained higher level of attractant, as observed for chemotaxis, does not seem to occur as the swimming speed remained at elevated levels for hours in \textit{R. sphaeoroides} and \textit{A. brasilense} \cite{Packer1994, Zhulin1993}. The discrepancy in the order of magnitude in the response times might be caused by the experimental set-ups, e.g. in \cite{Son2016}, the effector first needs to diffuse from two sides in a wide microfluidic channel before a uniform population response can be measured.
To conclude, further experiments could shed some light on the chemokinetic response function and adaptation time, which would benefit the further development of the model and its assumptions to understand how bacteria make use of chemokinesis in dynamic environments.

\section*{Data accessibility}
Code and data are available from the Zenodo repository: https://doi.org/10.5281/zenodo.4323421.

\section*{Acknowledgements}
We acknowledge financial support from EPSRC EP/L504920/1 and EP/N509620/1 (T.J.) and the Winton Programme for the Physics of Sustainability (T.J., O.A.C.), and funding from the European Union's Horizon 2020 research and innovation programme under the Marie Sk\l{}odowska-Curie grant agreement No 798411 (F.J.P.).

\section*{Authors' contributions}
TJ carried out analytical calculations and simulations. JL carried out initial calculations and simulations. FP assisted with initial simulations. TJ, FP and OC interpreted the simulations. 
TJ and OC wrote the manuscript, which was critically revised by FP. All authors gave final approval for publication and agree to be held accountable for the work performed therein.

\appendix
\section{Chemotactic drift and sensitivity}
\label{app:ChemotacticSensitivity}

Here we derive new relations for the chemotactic drift speed $V_\chi$ and sensitivity parameter $\chi$ for bacteria undergoing chemotaxis and chemokinesis in dynamic environments, connecting the macroscopic parameter $\chi$ with the microscopic swimming speed $v$ and tumbling rate $\alpha$. We follow the approach of de Gennes \cite{DeGennes2004, Locsei07, Croze2011}, considering run-and-tumble bacteria with a simplified {\it E. coli}-like chemotactic response. Our contribution for the purpose of this work is to include: i) a swimming speed that is a function of position; ii) a temporally varying contribution to the gradient perceived by the bacteria. As in \cite{DeGennes2004, Croze2011}, we consider a one-dimensional model of run-and-tumble bacteria propagating in gradient of attractant, to which bacteria are exhibiting a `small response', that is the tumbling response remains close to the adapted value \cite{Xue2015}. 
Chemotactic memory is modelled through kernel integral, reflecting the fact that bacteria `remember' their chemical environment over a characteristic delay time of a few seconds. We neglect directional persistence and rotational diffusion, which can be addressed in the same framework \cite{Locsei07}, but are here ignored for simplicity. 
The modified de Gennes model we employ could be applicable to chemokinetic species like rhizobia, which display run and tumble dynamics \cite{Armitage1997}. The model neglects features of the motility pattern of chemokinetic marine bacteria, such as a run-reverse-flick mechanism or the influence of a change in swimming speed on reorientation frequency, as observed for \textit{V. alginolyticus} \cite{Son2016}, but instead focus on the pure effect of speed change.


As the tumbling frequency follows an exponential distribution, tumbling events can be treated as independent events of a Poisson process with rate $\alpha(t)=\alpha_0 e^{-\Delta(t)}$, where $\alpha_0$ is the tumbling rate in the absence of a gradient and where $\Delta(t)$ is the chemotactic bias given by the memory integral
\begin{equation}
\Delta(t)=\int_{-\infty}^{t} dt' K(t-t') f_\chi (x(t')). \label{eq:chemobias}
\end{equation}
The concentration function was chosen as in the main text as $f_\chi=c(x,t)/(c(x,t)+k_\chi)$. The memory kernel $K(t)$ was first measured for \textit{E. coli} \cite{Segall} and more recently also for \textit{V. alginolyticus} \cite{Xie2015}. Based on these experiments we assume that the kernel obeys $\int_{0}^{\infty} K(t)dt=0$, i.e. the tumbling rate perfectly adapts. As mentioned, we shall assume a small response ($\Delta(t) \ll 1$), so that the bacterial tumble rate can be linearised to
\begin{equation}
\alpha(t)\approx \alpha_0 (1- \Delta (t)),
\label{eq:linearRate}
\end{equation}

Considering a run that starts at $t=0$, the probability density of a tumble event in the interval $[t,t+dt]$ is $\alpha(t) \exp(-\int_{0}^{t} dt' \alpha(t'))$. The mean run duration is given by
\begin{equation}
\tau = \left\langle \int_{0}^{\infty} \alpha(t) \exp \left( -\int_0^t dt' \alpha(t') \right) t \: dt \right\rangle_{paths},
\label{eq:MeanDuration1}
\end{equation}


where angled brackets denote averaging over all possible bacterial swimming paths, as the tumble rate is path dependent (the `paths' subscript will be omitted henceforth). Integrating (\ref{eq:MeanDuration1}) by parts and recalling (\ref{eq:linearRate}) we can write
\begin{equation}
\tau = \left\langle \int_{0}^{\infty} e^{-\alpha_0 t} \exp \left( \int_0^t dt' \Delta(t') \right) \: dt \right\rangle_{paths}.
\label{eq:MeanDuration1a}
\end{equation}
%
Next, since $\Delta(t) \ll 1$, we can linearise the exponential integral to obtain
\begin{equation}
\tau =\dfrac{1}{\alpha_0}+ \alpha_0 \int_{0}^{\infty}  dt e^{-\alpha_0 t}\left\langle \int_0^t dt' \Delta(t')\right\rangle,
\label{eq:MeanDuration2a}
\end{equation}
where we have brought the angled brackets inside the time integral to surround only path dependent quantities. Substituting the expression for the bias (\ref{eq:chemobias}) and performing a change of variables in the memory integral (\ref{eq:linearRate}) by defining $u=t-t'$, we then obtain
\begin{equation}
\begin{split}
&\tau = \dfrac{1}{\alpha_0}+ \alpha_0 \int_{0}^{\infty} dt e^{-\alpha_0 t}\times\\
&\left\langle \int_{0}^{t} dt' \int_{0}^{\infty} du \; K(u) \, f_\chi (x(t'-u)) \right\rangle.
\end{split}
\label{eq:MeanDuration2}
\end{equation}
%

Next, as the time interval of interest is small compared to the gradient variations, we can Taylor expand the concentration function about a reference position and time: 
\begin{equation}\label{fTaylor}
f_\chi(x(t-u))\approx \text{const}+x(t-u)\nabla f_\chi \mid_{t} +(t-u)\partial_tf_\chi \mid_{x(t)}.
\end{equation}
We note that the constant term does not influence the integral in \eqref{eq:MeanDuration2}, since the response function $K(u)$ integrates to zero. Further analysis is simplified by considering the special response function 
\begin{equation}\label{simpleresponse}
K(u)=A \delta(u-\theta),
\end{equation}
where $\theta$ is a single delay time \cite{DeGennes2004}. We note that this function does not represent a physical response and is considered solely for the purpose of facilitating the calculation. Then, substituting equations (\ref{simpleresponse}) and (\ref{fTaylor}) into \eqref{eq:MeanDuration2} and integrating over the delta function we obtain
%
%
%

\begin{equation}\label{eq:MeanDuration3}
\tau \approx \dfrac{1}{\alpha_0} + A \alpha_0  \int_{0}^{\infty} e^{-\alpha_0 t}  [I_1(t) + I_2(t)]  dt
\end{equation}
where we have defined
\begin{equation}
\begin{split}\label{I1}
I_1(t)&= \left\langle \int_{0}^{t} \nabla f_\chi x(t'- \theta)\,dt^\prime \right\rangle\\
&=  \nabla f_\chi \int_{0}^{t}  \left\langle x(t'- \theta)\right\rangle dt^\prime.
\end{split}
\end{equation}
The average over paths can be taken inside the time integral since tumbles for $t>0$ are treated as having no effect on cell motion, so that paths are independent of time on the interval considered. Proceeding similarly we define
\begin{equation}
I_2(t)=\int_0^t (t'- \theta) \left\langle\partial_tf_\chi \mid_{x(t')}\right\rangle\,dt^\prime.\label{I2}
\end{equation}


To evaluate (\ref{I1}), we recall that the bacteria we are considering do not undergo rotational diffusion and do not possess directional persistence. Thus, following \cite{DeGennes2004}, we see that for times preceding a run (when a bacterium is tumbling) the position of a bacterium is, when averaging over paths, not correlated to the velocity. On the other hand, during a run, the position correlates with velocity. In this way, in (\ref{I1}) we have
\begin{equation}
  \langle x(t'- \theta)\rangle =
    \begin{cases}
      0, & t'<\theta \\
      \pm v(x)(t'-\theta), \quad & t'>\theta\\
    \end{cases}       
\end{equation}
So that integral (\ref{I1}) becomes:
\begin{equation}\label{I11}
I_1(t)=\pm \nabla f_\chi v(x) \int_{0}^{t}(t'-\theta) \,dt^\prime =\pm \nabla f_\chi v(x) \frac{(t'-\theta)^2}{2} 
\end{equation}
We proceed analogously to evaluate integral (\ref{I2}). In this case, the integral over paths for times preceding a run does not average to zero, but sums up the contributions from the temporal variation of the gradient during tumbles. During a run, on the other hand, the temporal variation is evaluated for when bacteria are travelling up (down) the gradient, providing positive (negative) weights to the path integration, which provides
\begin{equation}
  \langle \partial_tf_\chi \mid_{x(t')}\rangle =
    \begin{cases}
      F(t',\theta), \quad & t'<\theta \\
     \pm\partial_tf_\chi,  & t'>\theta\\
    \end{cases}       
\end{equation}
where $F(t',\theta)$ is a function of time and delay. In this way, we can carry out integral (\ref{I2}) 
\begin{eqnarray}\label{I22}
I_2(t)&&=\int_0^\theta F(t',\theta) (t'- \theta)\,dt^\prime\\ \nonumber
&&\pm \partial_tf_\chi \mid_{x}\int_{
\theta}^{t}(t'-\theta) \,dt^\prime\\\nonumber 
&&=G(\theta, t) \pm \partial_tf_\chi \mid_{x} \frac{(t'- \theta)^2}{2} 
\end{eqnarray}
where we have defined $G(\theta, t)=\int_0^\theta F(t',\theta) (t'- \theta)\,dt^\prime$.  
Because it does not depend on direction up/down the gradient, this function (and its integrals) will cancel out in the evaluation of the run time, and will henceforth be omitted from the derivation. 
Inserting expressions (\ref{I11}) and (\ref{I22}) back into (\ref{eq:MeanDuration3}), we can thus write the mean run durations up ($R$) and down ($L$) the gradient as
\begin{equation}
\begin{split}
&\tau^{L,R} \approx \dfrac{1}{\alpha_0} + A \alpha_0 \int_{0}^{\infty} dt e^{-\alpha_0 t}\times\\
& \pm\left[ v(x)\nabla f_\chi + \partial_tf_\chi\right]  \frac{(t-\theta)^2}{2}, 
\end{split}
\end{equation}
which can be integrated to obtain
\begin{equation}
\tau^{L,R} \approx \dfrac{1}{\alpha_0} \pm \frac{v(x)}{\alpha_0^2} \left[\nabla f_\chi + \frac{1}{v(x)} \partial_tf_\chi \right]  A e^{-\alpha_0 \theta}.
\label{tauLRsingle}
\end{equation}
%
%
%
Extending to general response function $K(\theta)$ with a distribution of delays \cite{DeGennes2004}, equation (\ref{tauLRsingle})
\begin{equation}
\tau^{L,R} \approx \dfrac{1}{\alpha_0} \pm \frac{v(x)}{\alpha_0^2} \left[\nabla f_\chi + \frac{1}{v(x)} \partial_tf_\chi\right]  \int_{0}^{\infty} d\theta K(\theta) e^{-\alpha_0 \theta}.
\end{equation}
Since $\alpha_{L,R}=1/\tau_{L,R}$ and (see main text)
\begin{equation}
V_\chi= v(x) \frac{\alpha_L  - \alpha_R}{\alpha_R  + \alpha_L},
\end{equation}
we then arrive at an expression for the chemotactic drift
\begin{equation}
V_\chi = \dfrac{v(x)^2}{\alpha_0}\beta \left[\nabla f_\chi + \frac{1}{v(x)} \partial_tf_\chi\right], 
\end{equation}
where we have defined the constant $\beta=\int_{0}^{\infty} d\theta K(\theta) e^{-\alpha_0 \theta}$. 
The standard definition of the chemotactic sensitivity is obtained from the empirical drift $V_\chi=\chi \nabla f_\chi$, where as previously $\chi$ is the chemotactic sensitivity. We extend here the definition to include temporal gradients, and define
\begin{equation}
\chi(x)=\dfrac{v(x)^2}{\alpha_0}\beta,
\label{eq:chemSensitivity1}
\end{equation}
so that chemotactic drift is given by Eq.~\eqref{eq:chemotactic drift velocity Updated} in the main text.
In the absence of chemokinetic alterations to the swimming speed, the sensitivity is simply 
\begin{equation}
\chi_0 = \dfrac{v_0^2}{\alpha_0}\beta,
\label{eq:chemSensitivity2}
\end{equation}
where subscripts denote a constant swimming speed. Dividing \eqref{eq:chemSensitivity1} by \eqref{eq:chemSensitivity2}
provides relationship \eqref{eq: chemotactic sensitivity} in the main text.
%
%

\section{Numerical solution and parameter values}
\label{app: NumericalSolution}

A finite difference scheme was chosen to compute the numerical solution of the system of PDEs \eqref{eq: nondim model}. The domain $\Omega$ is represented by a vector of $M$ equally spaced grid points, $X_1,...,X_M$. 
A fourth order scheme in space and a first order scheme in time are used to approximate the derivatives.
Homogeneous Neumann boundary conditions are imposed for both substrate and bacterial concentration, i.e. $\frac{\partial C}{\partial X}|_{X=1,M}=0$ and $\frac{\partial B}{\partial X}|_{X=1,M}=0$.

In the case of steady attractant gradient (Fig.~\ref{fig:SteadyGradient} and Fig.~\ref{fig:SteadyGradientHill}) the initial attractant distribution is $C(X,0)=0.01X$, while the bacteria are uniformly distributed $B(X,0)=1.0$. For the self-generated gradient (Fig.~\ref{fig:AgarPlate} and Fig.~\ref{fig:AgarPlate_integration}) we assume a Gaussian bacterial inoculum, $B(R,0)=\exp(-R^2/\sigma^2)$, with a corresponding attractant distribution, $C(R,0)=1-\exp(-R^2/\sigma^2)$ \citep{Croze2011}. Finally, for Fig.~\ref{fig:PointSource} and Fig.~\ref{fig:PointSourceMaximum} the bacteria are initially uniformly distributed $B(R,0)=0.2$, while the attractant distribution is  $C(R,0)=S(4\pi N T)^{-1} \exp(-R^2/4T N)$, where $S=0.5$ and $T=0.02$.

The parameters used in the non-dimensional model are summarised in Table \ref{Tab: Parameters}. The parameters are mostly based on literature values for \textit{E. coli}. Due to the lack of experimental data required for a full estimate of the functional form of the chemokinetic response of, e.g., a single species, the values for the chemokinetic response ($\eta$, $\omega$, $n$) are motivated by several studies and chosen to illustrate the response. For example, $\eta=0.5$ corresponds to a $50\%$ maximum increase in the swimming speed, which is on the order of magnitude that has been reported for several species \cite{Packer1994, Deepika2015,Garren2014}. The largest increase reported, to the best of our knowledge, is $80\%$ for \textit{V. alginolyticus} \cite{Son2016}. 
Thus, $\eta=2$ is chosen as an extreme value to illustrate the effect of chemokinesis more clearly. 
The half-saturation constant $\omega$, which is required for the chemokinetic function given by Eq. \eqref{eq: nondim speed function}, has not been reported directly. Upon (visual) inspection of results, half-saturation constants may be about $0.1$mM of glucose and $0.1$mM of acetate for \textit{E.coli} \cite{Deepika2015} and \textit{R. sphaeroides} \cite{Packer1994}, respectively. In \cite{Garren2014}, the half maximum speed seems to be reached at about $30\%$ relative mucus concentration, while the maximum is reached at about $60\%$. As pointed out in \cite{Hein2016}, absolute concentration above a threshold can be detected before gradients can be accurately measured. Therefore, we assumed that the half-saturation constant of chemokinetic response ($\omega$) should be below or on the same order as the chemotactic half-saturation constant ($K_\chi$).
As there has not been a functional fit to chemokinesis measurements, an estimate for the Hill factor $n$ is difficult. From visual inspection of results in \cite{Garren2014} and \cite{Deepika2015} we may assume a factor between $1$ and $3$. Simulations with larger Hill factors were performed to compare to results based on the assumption of a step-change in swimming speed (i.e. $n \rightarrow \infty$) as done in agent-based models \cite{Garren2014, Son2016}.
The parameters used to produce the figures in the main text were chosen to illustrate the effect of chemokinesis best.
As a comparison, other parameter combinations are given in the Supplementary Material. 

\begin{table}[t]
\centering
\ra{1.3}
\caption{Parameters used in simulations. Please see main text for parameter definitions.}
\label{Tab: Parameters}
\begin{tabular}{lSlc} \toprule
Parameter& {Value} & Figure & Reference \\ \colrule
\multirow{2}{*}{$H$} & 0.0 & \ref{fig:SteadyGradient}, \ref{fig:SteadyGradientHill} & - \\ 
& 3.5 & \ref{fig:AgarPlate}, \ref{fig:PointSource}, \ref{fig:PointSourceMaximum} & \cite{Croze2011} \\ 
\colrule
\multirow{2}{*}{$N$} & 0.0 & \ref{fig:SteadyGradient}, \ref{fig:SteadyGradientHill} & - \\ 
& 0.5 & \ref{fig:AgarPlate}, \ref{fig:PointSource},  \ref{fig:PointSourceMaximum}& \cite{Croze2011} \\ 
\colrule
\multirow{2}{*}{$\delta_0$} & 50.0 & \ref{fig:SteadyGradient}, \ref{fig:SteadyGradientHill}, \ref{fig:PointSource} & - \\ 
& 105.0 & \ref{fig:AgarPlate} & \cite{Croze2011} \\ 
\colrule 
\multirow{2}{*}{$\omega$} & 0.2 & \ref{fig:SteadyGradient}, \ref{fig:SteadyGradientHill}, \ref{fig:PointSource} & - \\ 
& 0.5 & \ref{fig:AgarPlate} & - \\ 
\colrule 
\multirow{2}{*}{$\eta$} & 2.0 & \ref{fig:SteadyGradient}, \ref{fig:SteadyGradientHill}, \ref{fig:PointSource} & - \\ 
& 0.5 & \ref{fig:AgarPlate} & \cite{Garren2014} \\ 
\colrule
\multirow{4}{*}{$n$} & 1 & \ref{fig:SteadyGradientHill}, \ref{fig:PointSource} & - \\ 
& 5 & \ref{fig:SteadyGradient}, \ref{fig:SteadyGradientHill},\ref{fig:AgarPlate} & - \\ 
& 10 & \ref{fig:SteadyGradientHill} & - \\ 
& 40 & \ref{fig:SteadyGradientHill} & - \\ 
\colrule 
\multirow{2}{*}{$\zeta$} & 0.0 & \ref{fig:SteadyGradient}, \ref{fig:SteadyGradientHill},\ref{fig:AgarPlate} & - \\ 
& 8.164 {$\cdot 10^{-3}$} & \ref{fig:PointSource} & \cite{Croze2011} \\ 
\colrule
$K_S$ & 1.0 & \ref{fig:SteadyGradient}, \ref{fig:SteadyGradientHill},\ref{fig:AgarPlate}, \ref{fig:PointSource} & \cite{Croze2011} \\ 
\colrule
$K_\chi$ & 0.53 & \ref{fig:SteadyGradient}, \ref{fig:SteadyGradientHill},\ref{fig:AgarPlate}, \ref{fig:PointSource} & \cite{Croze2011} \\ 
\colrule
$S$ & 0.5 & \ref{fig:PointSource} \ref{fig:PointSourceMaximum} & \\
\botrule
\end{tabular} 
\end{table}



\setcounter{figure}{0}

\appendix*

\renewcommand{\thefigure}{S\arabic{figure}}
\renewcommand{\thetable}{S\arabic{table}}

\section*{SUPPLEMENTARY MATERIALS}

\subsection*{Additional simulations}

We present further simulation results for the scenarios described in the manuscript by varying crucial and unknown parameters determining the bacterial response. As the chemokinetic response function is unknown, these include the strength of the chemokinetic speed increase $\eta$, the half-saturation concentration of the chemokinetic response $\omega$ and the Hill factor $n$ as well as the strength of the chemotactic response $\delta_0$. 

\subsection*{Steady linear gradient}

\begin{figure}[tbph!]
	\centering
	\includegraphics*[width=\columnwidth]{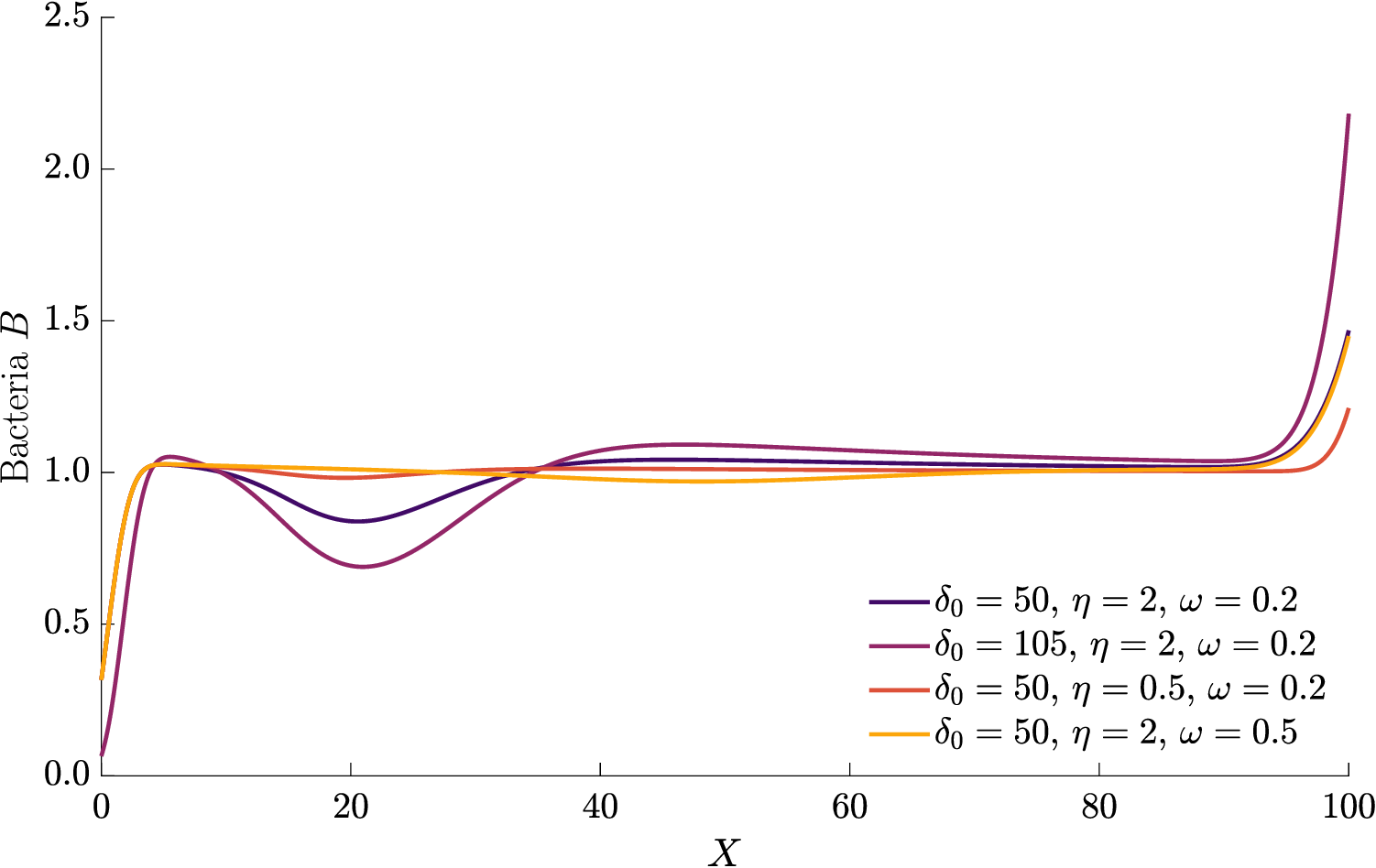}
	\caption{Bacterial response to a fixed linear attractant gradient as in Fig 3 in the main paper but with varying parameters ($T=1$).}
	\label{fig:SteadyGradient_Delta}
\end{figure}

The effect of modifying the chemotactic sensitivity $\delta_0$ and the chemokinetic parameters $\eta$ and $\omega$ is shown in Figure \ref{fig:SteadyGradient_Delta}.

\subsection*{Self-generated gradient}


The effect of varying the Hill parameter $n$ of the chemokinetic function $V(C)=1+\dfrac{\eta C^n}{C^n+\omega^n}$ is shown in Figure \ref{fig:AgarPlate}(a). Figure \ref{fig:AgarPlate}(b) illustrates the effect of the chemotactic sensitivity $\delta_0$ and the chemokinetic parameters $\omega$ and $\eta$.

\begin{figure}[tbph!]
	\centering
	\begin{subfigure}[b]{0.45\textwidth}
	\centering
	\includegraphics[width=\columnwidth]{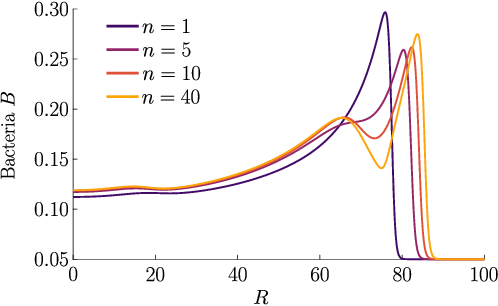}
	\caption{$T=17$}
	\label{fig:AgarPlateDelat0_pT171}	
	\end{subfigure} 
	\begin{subfigure}[b]{0.45\textwidth}
	\centering
	\includegraphics[width=\columnwidth]{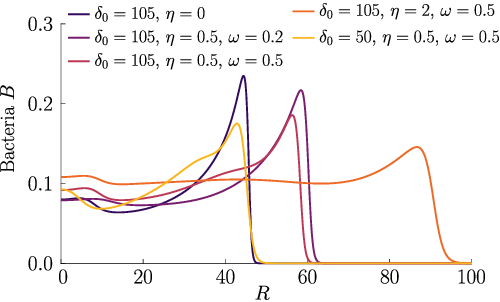}
	\caption{$T=11.9$}
	\label{fig:AgarPlateDelta0_pT120}
	\end{subfigure}
	\caption[Agar plate]{Bacterial response to a self-generated attractant gradient as in Fig. 5 in the main text but with varying parameters. (a) Varying the Hill parameter in the chemokinetic response function changes the bacterial density at $T=17$ (corresponds to the last column in Fig. 5 in the main text). (b) Varying chemotactic and chemokinetic parameters $\delta_0$, $\eta$ and $\omega$ at $T=11.9$, where a simulation using the parameters from the main paper is included as a comparison for this time step.}
	\label{fig:AgarPlate}
\end{figure}

\subsection*{Transient source}

The parameters used in the main text are the same as those used for the steady linear attractant gradient. The results of a simulation that uses the parameters of the self-generated attractant gradient instead are shown in Figure \ref{fig:PointSource}.

\begin{figure}[tbph!]
\includegraphics[width=\columnwidth]{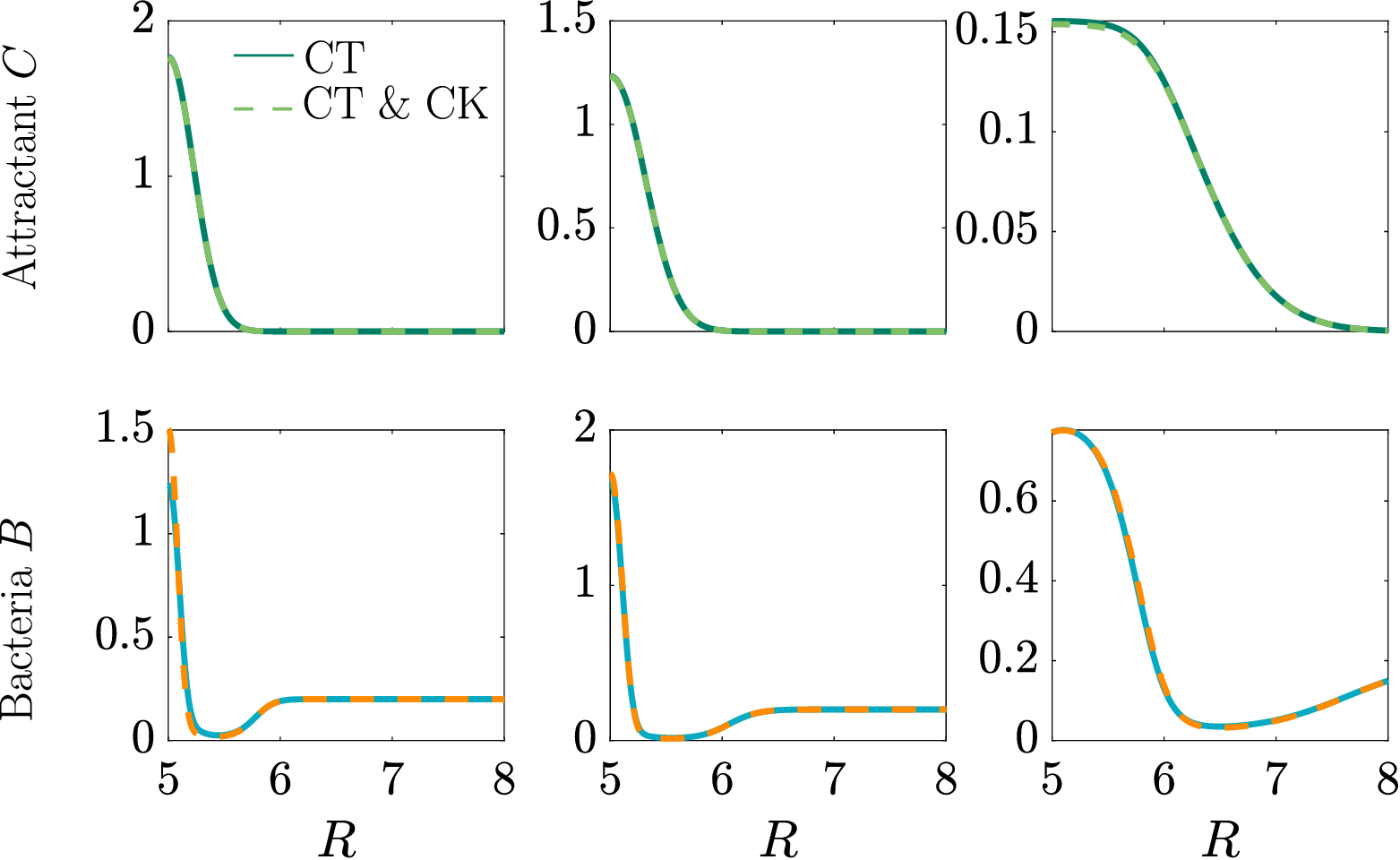}
\caption{Diffusing attractant from a transient source. Bacterial populations (bottom row) are attracted to source of diffusing attractant (top row). The chemokinetic-chemotactic population (orange curve) shows a faster and stronger accumulation than the purely chemotactic population (blue curve). Parameters $H=3.5, \, K_S=1, \, N=0.5, \, K_\chi=0.53, \, \delta_0=50,\, \eta=0.5, \, \omega=0.5, \, n=1, \, T=0.01,0.05,0.64$; no bacterial growth.}
	\label{fig:PointSource}
\end{figure}


\bibliographystyle{vancouver} 
\bibliography{references}

\end{document}